\documentclass[12pt,epsfig,color]{article}
\usepackage{amsmath}
\usepackage{epsfig}
\usepackage{amssymb}
\input{epsf}
\newcommand{\mysection}{\setcounter{equation}{0}\section}

\def\beq{\begin{equation}}
\def\eeq{\end{equation}}
\def\beqa{\begin{eqnarray}}
\def\eeqa{\end{eqnarray}}

\newcommand\as{\alpha_{\mathrm{S}}}
\newcommand\f[2]{\frac{#1}{#2}}

\def\beq{\begin{equation}}
\def\eeq{\end{equation}}
\def\beeq{\begin{eqnarray}}
\def\eeeq{\end{eqnarray}}
\def\to{\rightarrow}
\def\nn{\nonumber}

\def\b0{b_0}

\begin{document}

\begin{titlepage}
\renewcommand{\thefootnote}{\fnsymbol{footnote}}
\begin{flushright}
BNL-NT-07/28 \\
hep-ph/???
     \end{flushright}
\par \vspace{10mm}
\begin{center}
{\Large \bf Single-Inclusive hadron  production 
in polarized pp \\[5mm]  scattering at next-to-leading logarithmic accuracy}

\end{center}
\par \vspace{2mm}
\begin{center}
{\bf Daniel de Florian${}^{\,a}$,}
\hskip .2cm
{\bf Werner Vogelsang${}^{\,b}$,}
\hskip .2cm
and
\hskip .2cm
{\bf Federico Wagner${}^{\,a}$  }\\
\vspace{5mm}
${}^{a}\,$Departamento de F\'\i sica, FCEYN, Universidad de Buenos Aires,\\
(1428) Pabell\'on 1 Ciudad Universitaria, Capital Federal, Argentina\\
${}^{b}\,$Physics Department, 
Brookhaven National Laboratory, Upton, NY 11973, U.S.A.\\

\end{center}


\par \vspace{9mm}
\begin{center} {\large \bf Abstract} \end{center}
\begin{quote}
\pretolerance 10000
We study the resummation of large logarithmic perturbative 
corrections to the partonic cross sections relevant for the process 
$pp\to h X$ at high transverse momentum of the hadron $h$,
when the initial protons are longitudinally polarized. 
We perform the resummation to next-to-leading logarithmic accuracy.
We present numerical results for center-of-mass energies 
$\sqrt{S}=19.4$~GeV, relevant for comparisons to data from the 
Fermilab E704 experiment, and $\sqrt{S}=62.4$~GeV, where preliminary data
from RHIC have recently become available. We find significant enhancements of
the spin-dependent cross sections, but a decrease of the 
double-spin asymmetry for the process. This effect is less
pronounced at the higher energy. 
\end{quote}

\end{titlepage}

\setcounter{footnote}{1}
\renewcommand{\thefootnote}{\fnsymbol{footnote}}


\section{Introduction}

The spin structure of the nucleon continues to be a particular focus of
modern nuclear and particle physics. As is well known, the total quark and 
anti-quark (summed over all flavors) spin contribution to the nucleon spin 
was found to be only about $\sim 25 \%$, implying that the gluon spin 
contribution and/or orbital angular momenta may play an important role. 
There is currently much experimental activity aiming at further unraveling 
the nucleon's spin structure. One emphasis is on the determination of the 
spin-dependent gluon distribution, $\Delta g$,
of the nucleon, which ultimately would give the gluon spin contribution to
the nucleon spin. Deep-Inelastic scattering (DIS) has provided
most of the presently available information on nucleon spin structure, 
but has left $\Delta g$ essentially unconstrained~\cite{grsv, DS, deltag}. 
Particularly good prospects for determining $\Delta g(x,Q^2)$ over a 
wide range of momentum fractions $x$ and scales $Q$ are 
offered at the Relativistic Heavy-Ion Collider (RHIC) at BNL, which 
is the first polarized proton-proton collider. Spin asymmetries in 
high-energy $pp$ scattering can be particularly sensitive to $\Delta g$, 
for processes where gluons in the initial state contribute already at 
the lowest order of perturbation theory~\cite{Stratmann:2007hp}. One 
example is the single-inclusive production of large transverse-momentum 
($p_T$) hadrons, $pp \to h X$. Indeed, RHIC data taken 
at $\sqrt{S}=200$~GeV on the double-spin 
asymmetry $A_{LL}$ for $pp \to \pi X$~\cite{phenix} and for the related
process  $pp \to {\mathrm{jet}} X$~\cite{star} are now starting to put 
significant constraints on $\Delta g$, indicating that $\Delta g$ is 
not too sizable in the accessed region of gluon momentum fractions.
Similar conclusions are drawn from results obtained in lepton 
scattering~\cite{lepton}.

RHIC is, however, not the first place where the spin asymmetry $A^{\pi}_{LL}$ 
for $pp \to \pi X$ was investigated. The Fermilab E704 fixed-target 
experiment presented measurements of $A^{\pi}_{LL}$ 
for 200~GeV protons impeding 
on a proton target~\cite{experdata}, 
resulting in $\sqrt{S}=19.4$~GeV center-of-mass (c.m.) energy.
An asymmetry consistent with zero was found for pions produced with 
transverse momenta $1\leq p_T\leq 4$~GeV at central c.m. system angles. 
An interesting question is whether this information already puts a constraint 
on $\Delta g$ at the $x$ values relevant here, $0.1\lesssim x\lesssim 0.4$. 
In~\cite{experdata} the experimental data were also compared to theoretical 
leading-order (LO) calculations using various different $\Delta g$ 
distributions. It was found that indeed there was some sensitivity
of the data to $\Delta g$, with extremely large $\Delta g$ (of size similar
to the unpolarized gluon distribution in the accessed $x$ region) 
seemingly ruled out. On the other hand, 
there are arguments against such a direct interpretation. For typical
fixed-target kinematics as those in the E704 experiment, unpolarized 
single-inclusive hadron cross section data are generally not at all described
even by next-to-leading order (NLO) (let alone, LO) theoretical 
calculations~\cite{aur}, with theory falling way short. 
One may therefore wonder if it is then sensible
to confront LO calculations for $A^{\pi}_{LL}$ with the data. 
In order to address
this issue, Ref.~\cite{weber} considered the effects of possible 
Gaussian-distributed 
``intrinsic'' transverse momenta ($k_T$) of the partons on $A^{\pi}_{LL}$ 
in this kinematic regime.
It was found that intrinsic $k_T$ tends to decrease the spin asymmetry 
significantly, so that even relatively large $\Delta g$ appeared to be 
compatible with the E704 data. At the same time, intrinsic $k_T$ improves
the comparison with the unpolarized cross section data. However, from a
theoretical point of view, implementation of intrinsic $k_T$ into a 
single-inclusive cross section is not really a satisfactory approach because
one only has {\it collinear} factorization in this case. At best, intrinsic 
$k_T$ effects may be regarded as providing an effective model for possible
power-suppressed contributions to the cross section. Implementation of
intrinsic $k_T$ also obscures the role of perturbative higher-order
contributions to the cross section. Nonetheless, the results of~\cite{weber}
indicate that there can be substantial contributions to $A^{\pi}_{LL}$ 
in the fixed-target regime that go beyond low orders of perturbation theory.

Progress on the theoretical description of the unpolarized single-inclusive
hadron cross section in the fixed-target energy regime was made 
in Ref.~\cite{DW1}. For typical fixed-target kinematics, the value
of $x_T\equiv 2 p_T/\sqrt{S}$ is relatively large, $x_T\gtrsim 0.1$. 
It turns out that the partonic hard-scattering cross sections relevant
for $pp \to h X$ are then largely probed in the ``threshold''-regime, 
where the initial partons have just enough energy to produce the
high-transverse momentum parton that subsequently fragments into the
hadron, and its recoiling counterpart. Relatively little phase space 
is then available for additional radiation of partons. In particular,
gluon radiation is inhibited and mostly constrained to the emission
of soft and/or collinear gluons. The cancellation of infrared singularities 
between real and virtual diagrams then leaves behind large double- and
single-logarithmic corrections to the partonic cross sections. These
logarithms appear for the first time in the NLO expressions for the
partonic cross sections, where they arise as terms of the form 
$\as\ln^2(1-\hat{x}_T^2)$ in the rapidity-integrated cross section,
where $\hat{x}_T\equiv \hat{p}_T/\sqrt{\hat{s}}$ with $\hat{p}_T$
the transverse momentum of the produced parton and $\hat{s}$ the
c.m. energy of the initial partons. At yet higher ($k$th) order
of perturbation theory, the double-logarithms are of the form 
$\as^k\ln^{2k}(1-\hat{x}_T^2)$. When the threshold regime dominates,
it is essential to take into account the large logarithms to all 
orders in the strong coupling $\as$, a technique known as ``threshold
resummation''~\cite{dyresum}. Based on earlier work~\cite{KOS,Bon} on the 
resummation for $2\to 2$ QCD hard-scattering, we examined the
effects of threshold resummation on the single-inclusive
hadron cross section in~\cite{DW1} and found very significant
enhancements of the theoretical prediction in the fixed-target
regime, which in fact lead to a relatively good agreement between
resummed theory and the data. This also sheds light on the size of 
additional power-suppressed contributions to the cross sections 
(among them, perhaps, effects related to intrinsic $k_T$; see also
Ref.~\cite{bpr}), which do 
not seem to play a dominant role. We concluded that threshold resummation 
is an essential part of the theoretical description in the typical 
fixed-target kinematic regime. Its effects at higher energies (such as
at RHIC) are much smaller, even though it has to be said that one is 
here typically much further away from the threshold regime so that 
the applicability of threshold resummation is not entirely clear then. 

In the light of the results of Ref.~\cite{DW1} and the E704 data, 
it appears desirable to apply threshold resummation also to the 
spin asymmetry $A^{\pi}_{LL}$, which is the goal of this paper. In this
way, one may hope to put the theoretical description of $A^{\pi}_{LL}$
for single-inclusive hadron production in the fixed-target regime
on firmer ground. One may then also revisit the question as to whether
the E704 data already allow to put a constraint on $\Delta g$. 

We also note that recently preliminary 
data for the cross section and double-spin
asymmetry taken at RHIC's lower energy $\sqrt{S}=62.4$~GeV have been 
reported~\cite{Tann,experdata1}. Even though the approximations
needed for threshold resummation to be useful work slightly worse 
for the kinematics relevant here, it is of great interest to confront the
resummation with the data. This will also be done in this paper.

The remainder of this paper is organized as follows: Section~\ref{sec2} 
summarizes the theoretical perturbative-QCD framework for the process 
under study. In Sec.~\ref{sec3}, we present the resummed spin-dependent 
cross section to next-to-leading logarithmic (NLL) accuracy.  
Section~\ref{sec4} gives phenomenological results for the effects
of threshold resummation on the spin-dependent high-$p_T$ pion cross section 
at $\sqrt{S}=19.4$~GeV and at $\sqrt{S}=62.4$~GeV, 
and on the corresponding double-spin asymmetries 
$A^{\pi}_{LL}$. Finally we draw our conclusions in Sec.~\ref{sec5}. 
Two Appendices contains the relevant ingredients for the resummation in the
spin-dependent case.


\section{Cross section and spin asymmetry in perturbation \\
theory
\label{sec2}}

We are considering the process 
\begin{align}
p(p_1,\Lambda_1) + p(p_2, \Lambda_2) \rightarrow h(p_3) + X \, ,
\end{align}
where the $\Lambda_i$ denote the helicities of the initial protons, and
the $p_i$ ($i=1,2,3$) are the four-momenta of the ``observed'' hadrons.
One defines the spin-averaged and spin-dependent cross sections
as
\beqa
d\sigma &=&\frac{1}{2}\left[ 
d\sigma(\Lambda_1=+,\Lambda_2=+)+
d\sigma(\Lambda_1=+,\Lambda_2=-)\right] \; , \nonumber \\
d\Delta \sigma &=&\frac{1}{2}\left[ 
d\sigma(\Lambda_1=+,\Lambda_2=+)-
d\sigma(\Lambda_1=+,\Lambda_2=-)\right] \; , \label{pdef}
\eeqa
respectively, and their double-spin asymmetry as
\beq
A_{LL} =\frac{d\Delta \sigma}{d\sigma} \; . \label{alldef}
\eeq
Hadron $h$ is assumed to be produced at 
large transverse momentum $p_T$. For such a large-momentum-transfer
reaction, the factorization theorem~\cite{FT} states that cross section 
may be factorized in terms of collinear convolutions of parton
distribution functions for the initial protons, a fragmentation 
function for the final-state hadron, and short-distance parts
that describe the hard interactions of the partons and are amenable
to QCD perturbation theory. The long-distance parton distributions and
fragmentation functions are universal, i.e., they are the same in any 
inelastic reaction. Long- and short-distance contributions are 
separated by a factorization scale. 

As discussed in Ref.~\cite{DW1}, a major simplification of the resummation 
formalism occurs when the cross section is integrated over all
pseudo-rapidities $\eta$ of the produced pion. This will also be
done in this paper. The factorized spin-dependent cross section for 
$pp\to hX$ can then be written as
\begin{align}
\label{eq:1} \f{p_T^3\, d\Delta\sigma(x_T)}{dp_T} = \sum_{a,b,c}\, &
\int_0^1 dx_1 \, \Delta f_a\left(x_1,\mu^2\right) \, \int_0^1
dx_2 \, \Delta f_b\left(x_2,\mu^2\right) \, \int_0^1 dz
\,z^2\, D_{h/c}\left(z,\mu^2\right) \, \nn \\ &\times \int_0^1
d\hat{x}_T \, \, \delta\left(\hat{x}_T-\f{x_T}{z\sqrt{x_1
x_2}}\right) \, \int_{\hat{\eta}_{-}}^{\hat{\eta}_{+}} d\hat{\eta}
\, \f{\hat{x}_T^4 \,\hat{s}}{2} \,
\f{d\Delta \hat{\sigma}_{ab\rightarrow cX}
(\hat{x}_T^2,\hat{\eta})}{ d\hat{x}_T^2 d\hat{\eta}} \, .
\end{align}
Here the $\Delta f_{a,b}$ are the spin-dependent parton distributions
in the proton, 
\beq
\Delta f_a(x,\mu^2)=f_a^+(x,\mu^2)-f_a^-(x,\mu^2)
\eeq
with $f_a^+$ ($f_a^-$) denoting the distribution of parton type $a$
with positive (negative) helicity in a proton of positive helicity.
$D_{h/c}$ is the fragmentation function for parton $c$ fragmenting
into the observed high-$p_T$ hadron. The sum in Eq.~(\ref{eq:1})
runs over all partonic channels, with the associated
spin-dependent partonic cross sections $d\Delta 
\hat{\sigma}_{ab\rightarrow cX}$. The latter are defined analogously
to Eq.~(\ref{pdef}), with helicities now corresponding to parton
ones. They are perturbative and have an expansion of the form
\beq
d\Delta 
\hat{\sigma}_{ab\rightarrow cX}=d\Delta 
\hat{\sigma}_{ab\rightarrow cX}^{(0)}+\frac{\alpha_s}{\pi}d\Delta 
\hat{\sigma}_{ab\rightarrow cX}^{(1)}+\ldots
\eeq
with $\as$ the strong coupling. 
The scale $\mu$ in Eq.~(\ref{eq:1}) 
is the factorization scale. We could distinguish
in principle between factorization scales for the initial state
(parton distributions) and the final state (fragmentation function).
For simplicity, we will not do this in this paper. There is also
a renormalization scale, at which the strong coupling constant 
is evaluated. We will collectively denote all scales by $\mu$.
The dependence on $\mu$ is implicit in the partonic cross sections in
Eq.~(\ref{eq:1}). Finally, $\hat{\eta}$ is the pion's pseudorapidity at 
parton level, related to the one at hadron level by $\hat{\eta}=\eta 
-\frac{1}{2}\ln(x_1/x_2)$. Its limits are given by
$\hat{\eta}_{+}=-\hat{\eta}_{-}=\ln\left[(1+\sqrt{1-\hat{x}_T^2})/
\hat{x}_T\right]$ where, as before, $x_T\equiv 2 p_T/\sqrt{S}$, and
its partonic counterpart is $\hat{x}_T\equiv 2 p_T^c/\sqrt{\hat{s}}=
x_T/z\sqrt{x_1x_2}$.

We note that the corresponding expression for the factorized
spin-averaged cross section is obtained from Eq.~(\ref{eq:1}) by
dropping all $\Delta$'s, meaning that the spin-dependent parton
distributions are replaced by their usual unpolarized counterparts, 
and the partonic scattering cross sections by the spin-averaged ones.

\section{Resummed cross section \label{sec3}}

As mentioned above, we will follow~\cite{DW1} to perform the threshold
resummation only for the case of the fully rapidity-integrated cross section.
The resummation of the soft-gluon contributions is achieved by taking a
Mellin transform of the cross section in the scaling variable $x_T^2$:
\begin{align}
\label{eq:moments}
\Delta \sigma(N)\equiv \int_0^1 dx_T^2 \, \left(x_T^2 \right)^{N-1} \;
\f{p_T^3\, d\Delta \sigma(x_T)}{dp_T} \, .
\end{align}
For the rapidity-integrated cross section, the convolutions in 
Eq.~(\ref{eq:1}) between parton distributions, fragmentation functions, 
and subprocess cross sections then become ordinary products~\cite{DW1,CMN}:
\begin{align}
\Delta \sigma(N)=\sum_{a,b,c} \,  \Delta f_a(N+1,\mu^2) \,
\Delta f_b(N+1,\mu^2) \,  D_{h/c}(2N+3,\mu^2) \,
\Delta \hat{\sigma}_{ab\rightarrow cX}(N)\, ,
\end{align}
where
\begin{align} \label{momdef}
\Delta \hat{\sigma}_{ab\rightarrow cX}(N) \equiv
\int_0^1 d\hat{x}_T^2 \, \left(\hat{x}_T^2 \right)^{N-1}\,
\int_{\hat{\eta}_{-}}^{\hat{\eta}_{+}} d\hat{\eta}
\, \f{\hat{x}_T^4 \,\hat{s}}{2} \,
\f{d\Delta \hat{\sigma}_{ab\rightarrow cX}
(\hat{x}_T^2,\hat{\eta})}{ d\hat{x}_T^2 d\hat{\eta}} \,.
\end{align}
In Mellin-moment space, the threshold logarithms become logarithms
in the moment variable $N$. The leading logarithms are of the form
$\as^k \ln^{2k}N$; subleading ones are down by one or more powers
of $\ln N$. Threshold resummation results in exponentiation of 
the soft-gluon corrections in moment space~\cite{dyresum,KOS}. 
The leading logarithms are contained in radiative factors for the 
initial and final partons. Because of color interferences and 
correlations in large-angle soft-gluon emission at NLL, for QCD
hard-scattering the resummed cross section becomes a sum of 
exponentials, rather than a single one~\cite{DW1,KOS}, unlike the
much simpler cases of the Drell-Yan or Higgs cross sections~\cite{dyresum}.

Combining results of~\cite{dyresum,KOS,cc}, we can cast the resummed 
spin-dependent partonic cross section for each subprocess 
into a relatively simple form~\cite{DW1}\footnote{Note that the 
symbols ${\cal D}_N^i, {\cal D}^{{\rm (int)} ab\rightarrow cd}_{I\, N}$ 
in the equation below are usually referred to as 
$\Delta_N^i,\Delta^{{\rm (int)} ab\rightarrow cd}_{I\, N}$ 
in the literature~\cite{DW1}. We have changed this notation
in order to avoid confusion with the label ``$\Delta$'' indicating
spin-dependent cross sections and parton distributions in this paper.}:
\begin{align}
\label{eq:res}
\Delta \hat{\sigma}^{{\rm (res)}}_{ab\to cd} (N)= \Delta C_{ab\to cd}\,
{\cal D}^a_N\, {\cal D}^{b}_N\, {\cal D}^{c}_N\,
J^{d}_N\, \left[ \sum_{I} \Delta G^{I}_{ab\to cd}\,
{\cal D}^{{\rm (int)} ab\rightarrow cd}_{I\, N}\right] \,
\Delta \hat{\sigma}^{{\rm (Born)}}_{ab\to cd} (N) \;  ,
\end{align}
where $\Delta \hat{\sigma}^{{\rm (Born)}}_{ab\to cd}(N)$ denotes the 
LO term in the perturbative expansion of Eq.~(\ref{momdef}) for each 
process. We list the moment-space expressions for all the spin-dependent
Born cross sections in Appendix A. Each of the functions  
$J^{d}_N$,${\cal D}^{i}_N$,${\cal D}^{{\rm (int)} ab\rightarrow cd}_{I\, N}$
in Eq.~(\ref{eq:res}) is an exponential. ${\cal D}^a_N$ represents the 
effects of soft-gluon radiation collinear to initial parton $a$ and is
given, in the $\overline{{\mathrm{MS}}}$ scheme, by
\begin{align}\label{Dfct}
\ln {\cal D}^a_N&=  \int_0^1 \f{z^{N-1}-1}{1-z}
\int_{\mu^2}^{(1-z)^2 Q^2} \f{dq^2}{q^2} A_a(\as(q^2)) dz\; ,
\end{align}
and similarly for $\Delta^b_N$. Here, $Q^2=2 p_T^2$. We will 
specify the function $A_a$ below. Collinear soft-gluon
radiation to parton $c$ yields the same function~\cite{cc}. 
The function $J^{d}_N$ embodies collinear, soft or hard, emission 
by the non-observed recoiling parton $d$ and reads:
\begin{align} \label{Jfct}
\ln J^d_N&=  \int_0^1 \f{z^{N-1}-1}{1-z} \Big[
\int_{(1-z)^2 Q^2}^{(1-z) Q^2} \f{dq^2}{q^2} A_a(\as(q^2)) +
\f{1}{2} B_a(\as((1-z)Q^2)) \Big] dz\; .
\end{align}
Large-angle soft-gluon emission is accounted for by the factors
${\cal D}^{{\rm (int)} ab\rightarrow cd}_{I\, N}$, which depend on
the color configuration $I$ of the participating partons. A sum over
the latter occurs in Eq.~(\ref{eq:res}), with $\Delta G^{I}_{ab\to cd}$ 
representing a weight for each color configuration, 
such that $\sum_I \Delta G^{I}_{ab\to cd}=1$. Each of the 
${\cal D}^{{\rm (int)} ab\rightarrow cd}_{I\, N}$ is given as
\begin{align}\label{Dintfct}
\ln{\cal D}^{{\rm (int)} ab\rightarrow cd}_{I\, N} &=
 \int_0^1 \f{z^{N-1}-1}{1-z} D_{I\, ab\to cd}(\as((1-z)^2 Q^2))dz \; .
\end{align}
Finally, the coefficients $\Delta C_{ab\to cd}$ contain
$N-$independent hard contributions arising from one-loop
virtual corrections.

Most of the ingredients to Eqs.~(\ref{Dfct})-(\ref{Dintfct}) are
well-known from the literature because they coincide with the 
results obtained for spin-averaged scattering. This is the case
for the functions $A_a$, $B_a$, and $D_{I\, ab\to cd}$, because these
are associated with soft gluon emission, which is spin-independent.
The only differences between the spin-dependent and the spin-averaged
cases reside in the coefficients $\Delta G^{I}_{ab\to cd},
\Delta C_{ab\to cd}$ and of course
in the Born cross sections. These terms are all related to hard
radiation, which depends on the polarization state and therefore in 
general differs for the polarized and unpolarized cases.   

We first briefly recall the known 
functions and then turn to the new parts. 
Each of the functions ${\cal F}\equiv A_a$, $B_a$, $D_{I\, ab\to cd}$
is a perturbative series in $\as$,
\begin{equation}
{\cal F}(\as)=\frac{\as}{\pi} {\cal F}^{(1)} +
\left( \frac{\as}{\pi}\right)^2 {\cal F}^{(2)} + \ldots \; ,
\end{equation}
with~\cite{KT}:
\begin{equation}
\label{A12coef}
A_a^{(1)}= C_a
\;,\;\;\;\; A_a^{(2)}=\frac{1}{2} \; C_a  \left[
C_A \left( \frac{67}{18} - \frac{\pi^2}{6} \right)
- \frac{5}{9} N_f \right]
\;,\;\;\;\; B_a^{(1)}=\gamma_a \; ,
\end{equation}
where $N_f$ is the number of flavors, and
\begin{eqnarray}
&&C_g=C_A=N_c=3 \;, \;\;\;C_q=C_F=(N_c^2-1)/2N_c=4/3 \nn \\
&&\gamma_q=-3 C_F/2=-2\; , \;\;\; \gamma_g=-2\pi \b0\; , \;\;\;
\b0 = \frac{1}{12 \pi} \left( 11 C_A - 2 N_f \right) \; .
\end{eqnarray}
The coefficients $D_{I\, ab \to c d}^{(1)}$ are listed in Ref.~\cite{DW1}.

In order to determine the effects of the color interferences for large-angle 
soft-gluon emission in the $2\to 2$ processes $ab\to cd$, we can follow
the procedures presented in~\cite{KOS,Bon}. The soft-anomalous dimensions
and soft factors determined in~\cite{KOS} are again identical in the 
spin-averaged and the spin-dependent cases. The differences arise
solely in the color-connected Born cross sections. We therefore
only need to derive the latter for polarized scattering, using
the same color basis as that chosen in~\cite{KOS}. The results 
in~\cite{KOS} have actually been given for arbitrary rapidity; for the
case of the rapidity-integrated cross section we consider here it 
is sufficient to set $\hat{\eta}=0$ (see \cite{DW1} for more detail).
Nonetheless, for future convenience, we present in this work the 
spin-dependent color-connected Born cross sections also
at arbitrary rapidity. In this
way, they may be directly used in future investigations of the 
resummed cross section at fixed rapidity. The results are collected in
Appendix~B. For the case of the rapidity-integrated cross section,
the color-connected Born cross sections at $\hat{\eta}=0$, when normalized
to the full Born cross section for each partonic channel, give
the color weights $\Delta G^{I}_{ab\to cd}$~\cite{DW1} that we need to NLL,
which are listed in Appendix~A.

The perturbative expansion of the coefficients $\Delta C_{ab\to cd}$ reads:
\begin{eqnarray}
\Delta C_{ab\to cd} = 1 + \frac{\as}{\pi} \Delta C_{ab\to cd}^{(1)} + 
{\cal O}(\as^2) \; .
\end{eqnarray}
In order to determine the coefficients $\Delta C_{ab\to cd}^{(1)}$, we take
advantage of the full analytic NLO calculation of Ref.~\cite{Jager}. For
each partonic channel one expands the resummed cross section in 
Eq.~(\ref{eq:res}) to first order in $\as$. Near threshold, one
can straightforwardly take Mellin moments of the full NLO expressions
of~\cite{Jager}. By comparison of the two results one first verifies
that all logarithmic terms in the full NLO results are correctly 
reproduced by the resummation formalism. The remaining $N$-independent
terms in the NLO cross section give the coefficients 
$\Delta C_{ab\to cd}^{(1)}$. These turn out to have rather lengthy 
expressions, and we only give them in numerical form in Appendix~A.

This completes the collection of the ingredients for the resummed
partonic cross sections. In the exponents, the large logarithms in 
$N$ occur only as {\it single} logarithms, of the form
$\as^k \ln^{k+1}(N)$ for the leading terms. Next-to-leading logarithms
are of the form $\as^k \ln^k(N)$. Knowledge of the
coefficients given above allows to resum the full LL and NLL full towers
in the exponent. It is useful to expand the resummed exponents 
to definite logarithmic order:
\beeq \label{lndeltams} \!\!\! \!\!\! \!\!\! \!\!\!
\!\!\! \ln {\cal D}_N^a(\as(\mu^2),Q^2/\mu^2)
&\!\!=\!\!& \ln N \;h_a^{(1)}(\lambda) +
h_a^{(2)}(\lambda,Q^2/\mu^2) + {\cal O}\left(\as(\as
\ln N)^k\right) \,,\\ \label{lnjfun} \ln
J_N^a(\as(\mu^2),Q^2/\mu^2) &\!\!=\!\!& \ln N \;
f_a^{(1)}(\lambda) + f_a^{(2)}(\lambda,Q^2/\mu^2) + {\cal
O}\left(\as(\as \ln N)^k\right) \; , \eeeq 
where $\lambda=\b0 \as(\mu^2) \ln N$. The functions $h^{(i)}$ and
$f^{(i)}$ have for example been given in~\cite{DW1}. $h^{(1)}$ and $f^{(1)}$ 
contain all LL terms in the perturbative series, while $h^{(2)}$ and $f^{(2)}$
are only of NLL accuracy. For a complete NLL resummation one also
needs the coefficients $\ln {\cal D}^{{\rm (int)} ab\rightarrow
cd}_{I\, N}$ whose NLL expansion reads: 
\beeq \label{lndeltams1}
\ln{\cal D}^{{\rm (int)} ab\rightarrow cd}_{I\,
N}(\as(\mu_R^2),Q^2/\mu_R^2) &\!\!=\!\!& \frac{D_{I\, ab \to c
d}^{(1)}}{2\pi b_0} \;\ln (1-2\lambda) + {\cal O}\left(\as(\as \ln
N)^k\right) \, . 
\eeeq 

In order to obtain a resummed cross section in $x_T^2$ space, one needs 
an inverse Mellin transform. Here one has to deal with the singularity
in the perturbative strong coupling constant in 
Eqs.~(\ref{Dfct})-(\ref{Dintfct}), which manifests itself also in the 
singularities of the functions $h^{(1,2)}$ and $f^{(1,2)}$ above at 
$\lambda=1/2$ and $\lambda=1$. We use the {\em Minimal Prescription} 
developed in Ref.~\cite{Catani:1996yz}, which relies on use of the 
NLL expanded forms Eqs.~(\ref{lndeltams})-(\ref{lndeltams1}), 
and on choosing a Mellin contour in complex-$N$ space that 
lies to the {\it left} of the poles at $\lambda=1/2$ and $\lambda=1$ 
in the Mellin integrand:
\begin{align}
\label{hadnmin}
\f{p_T^3\, d\Delta \sigma^{\rm (res)}(x_T)}{dp_T} &=
\;\int_{C_{MP}-i\infty}^{C_{MP}+i\infty}
\;\frac{dN}{2\pi i} \;\left( x_T^2 \right)^{-N}
\Delta \sigma^{\rm (res)}(N) \; ,
\end{align}
where $b_0\as(\mu^2)\ln C_{MP}<1/2$, but all other poles
in the integrand are as usual to the left of the contour. The
result defined by the Minimal Prescription has the property that
its perturbative expansion is an asymptotic series that
has no factorial divergence and therefore
no ``built-in'' power-like ambiguities. 

Finally, in order to make full use of the available fixed-order 
cross section~\cite{Jager, DdF}, which in our case is NLO (${\cal O}(\as^3)$), 
we match the resummed cross section to the NLO one. 
We expand the resummed cross section to ${\cal O}(\as^3)$, subtract the 
expanded result from the resummed one, and add the full NLO cross section:
\begin{align}
\label{hadnres}
\f{p_T^3\, d\Delta \sigma^{\rm (match)}(x_T)}{dp_T} &= \sum_{a,b,c}\,
\;\int_{C_{MP}-i\infty}^{C_{MP}+i\infty}
\;\frac{dN}{2\pi i} \;\left( x_T^2 \right)^{-N+1}
\; \Delta f_{a/P_1}(N,\mu^2) \; \Delta f_{b/P_2}(N,\mu^2) \;
D_{c/h}(2N+1,\mu^2)
 \nn \\
&\times \left[ \;
\Delta \hat{\sigma}^{\rm (res)}_{ab\to cd} (N)
- \left. \Delta \hat{\sigma}^{{\rm (res)}}_{ab\to cd} (N)
\right|_{{\cal O}(\as^3)} \, \right]
+\f{p_T^3\, d\Delta \sigma^{\rm (NLO)}(x_T)}{dp_T}
 \;\;,
\end{align}
where $\Delta \hat{\sigma}^{{\rm (res)}}_{ab\to cd} (N)$ is the 
polarized resummed cross section for the partonic channel $ab\to cd$ 
as given in Eq.~(\ref{eq:res}). In this way, NLO is taken into account 
in full, and the soft-gluon contributions beyond NLO are resummed to NLL. 
Any double-counting of perturbative orders is avoided.


\section{Phenomenological Results \label{sec4}}

We are now in the position to present numerical results for the 
threshold-resummed spin-dependent cross section and spin asymmetry in 
single-inclusive pion production in hadronic collisions. We will focus 
here on $pp\to \pi^0 X$ in fixed-target scattering at $\sqrt{S}=19.4$~GeV 
and at $\sqrt{S}=62.4$~GeV at the RHIC collider. In both these cases, 
experimental data exist~\cite{experdata,experdata1}. 

For our calculations, we need to choose sets of parton 
distribution and pion fragmentation functions. To study the 
sensitivity of the measured spin asymmetries to the spin-dependent
parton distribution functions, in particular the gluon density 
$\Delta g$, we use the ``Gl\"{u}ck-Reya-Stratmann-Vogelsang 
(GRSV)''~\cite{grsv} and the ``de Florian-Sassot (DS)''~\cite{DS} 
densities. These both offer various sets of distributions, distinguished 
mostly by $\Delta g$ distributions of different sizes. 
The set labeled ``GRSV'' is the regular (``standard'') GRSV set. We will
also use the GRSV ``max G'' set, which has a much larger $\Delta g$,
given by assuming $\Delta g(x)=g(x)$ at the initial scale for the
parton evolution. Likewise, the DS ``i+'' was constrained to have 
a much lower $\Delta g$
than the ``iii+'' one. For the spin-averaged cross section, we employ the
MRST2002~\cite{mrst} set throughout. The pion fragmentation functions 
are taken from the most recent analysis of $e^+e^-$ and $pp$ data, 
``de Florian-Sassot-Stratmann (fDSS)'' Ref.~\cite{deFlorian:2007aj}.
It is worth noticing that, according to Eq.~(\ref{hadnres}), one would like
to have the parton densities and fragmentation functions
available in Mellin-moment space. Technically, since most of the 
distributions are only available in $x$ space, we first perform a fit 
with a simple functional form to the distributions, of which we are 
able to take moments analytically. This has to be done separately
for each parton type and at each factorization scale. 

\begin {figure}[t]
\begin{center}
\includegraphics[width = 3.5in]{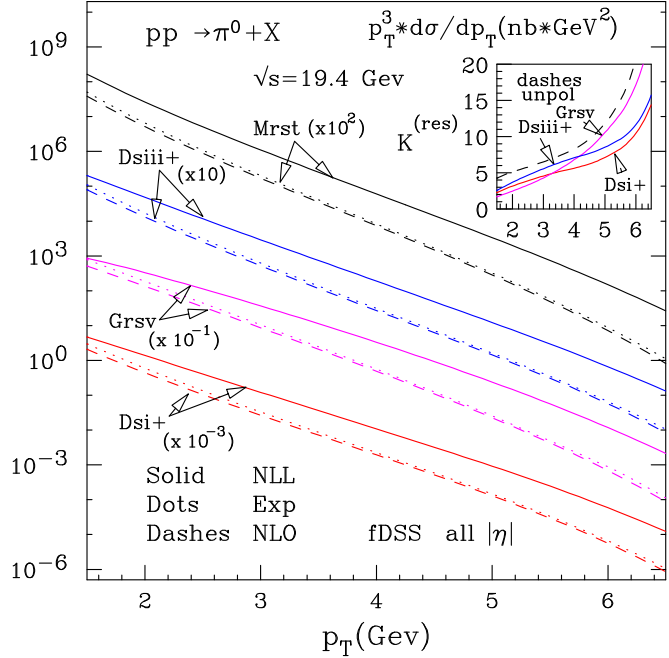}  
\end{center}
\caption{{\it NLO and NLL resummed cross sections for polarized 
$pp \to \pi^0 X $ at $\sqrt{S}=19.4$~GeV, for various sets of 
spin-dependent parton distributions of~\cite{grsv,DS}. We also
show the ${\cal O}(\as^3)$ expansions of the resummed cross sections,
and the analogous results in the unpolarized case. For better 
visibility, we have applied numerical factors to some results, 
as indicated in the figure. In the upper right inset, we present the 
ratios between the NLL resummed cross sections and the NLO ones.}
\label{figres315} }
\end{figure}

In Fig.~\ref{figres315} we present our results for the spin-dependent
cross section at $\sqrt{S}= 19.4$~GeV, integrated over all pion 
pseudo-rapidities $\eta$, for the various sets of the 
polarized parton distributions of~\cite{grsv,DS}. We have chosen 
all scales as $\mu=p_T$. We show the full NLO cross
section based on the calculations in~\cite{Jager}, as well as 
the NLL resummed predictions. We also display the expansions of the
resummed cross section to ${\cal O}(\as^3)$, which is the first order
beyond LO. As can be observed, these
faithfully reproduce the full NLO result, implying that the threshold
logarithms addressed by resummation indeed dominate the cross section
in this kinematic regime, so that resummation is expected to be
useful. Towards lower $p_T\sim 3$ GeV, the expansions slightly 
overestimate the NLO cross section, which is expected since one 
is further away from the threshold regime here, so that the
soft-gluon approximation tends to become less reliable.   
For the sake of completeness the unpolarized case is also represented 
in Fig.~\ref{figres315}.

The inset in Fig.~\ref{figres315} shows the resummed ``$K$-factors''
for the cross sections, defined as the ratios of the
resummed cross sections to the NLO cross sections (polarized or unpolarized) :
\begin{equation}
\label{eq:kres}
K^{{\rm (res)}} = \f{{d\sigma^{\rm (match)}}/{dp_T}}
{{d\sigma^{\rm (NLO)}}/{dp_T}}\, .
\end{equation}
As can be seen, $K^{{\rm (res)}}$ is very large, meaning that resummation 
results in a dramatic enhancement over NLO. For the unpolarized case,
this finding is in line with our previous results in~\cite{DW1}. It 
is interesting to see that $K^{{\rm (res)}}$ is large also for all sets of
spin-dependent parton distributions. It is evident, however, that the
enhancement is somewhat smaller than in the unpolarized case. This immediately
implies that the spin asymmetry $A_{LL}^{\pi}$ will generally be reduced 
when going from NLO to the NLL resummed case. One also notices that the 
resummation effects vary slightly for the various polarized parton densities.
This may be understood from the fact that $\Delta g$ is of different
size in the various sets. Typically, resummation effects are more
important for partonic channels with more external gluons~\cite{DW1}, 
so the size of $\Delta g$ matters. 

\begin {figure}[t]
\begin{center}
\includegraphics[width = 3.5in]{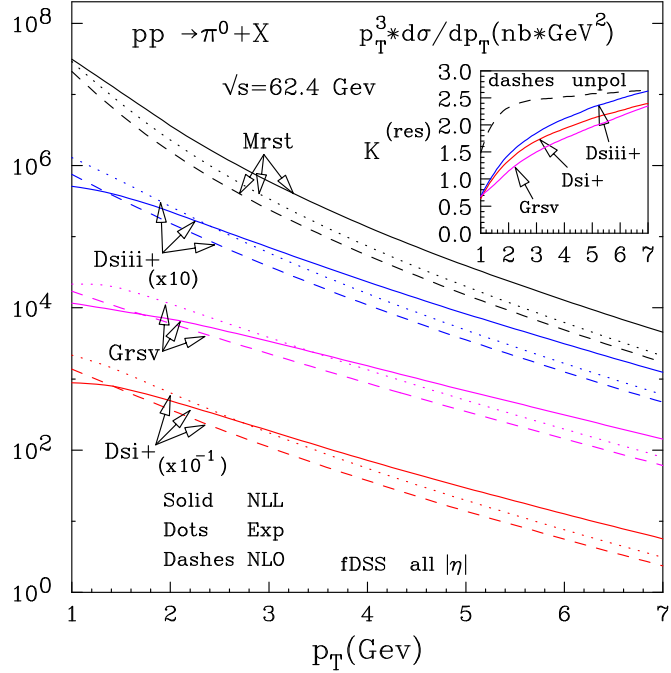}  
\end{center}
\caption{{\it Same as Fig.~\ref{figres315} but for $\sqrt{S}=62.4$ GeV.}
\label{figres2}}
\end{figure}

Fig.~\ref{figres2} shows similar results for $pp\to \pi^0 X$ at $\sqrt{S}= 
62.4$~GeV. As expected from the fact that one is further away from
threshold here, the soft-gluon approximation becomes somewhat less accurate
in this case, in particular at the lower $p_T$. 
One can observe that the resummation effects 
are generally much smaller at $\sqrt{S}= 62.4$~GeV than at 
$\sqrt{S}=19.4$~GeV. 

As mentioned earlier, we have determined the resummed formulas
for the fully rapidity-integrated cross section, whereas in experiments 
typically only a certain limited range in $\eta$ 
is covered. In order to be able
to compare to data, we therefore approximate the cross section 
(polarized or unpolarized) in the experimentally accessible rapidity region 
by
\begin{equation}
\f{p_T^3\, d\sigma^{\rm (match)}}{dp_T}({\rm \eta\, in\, exp.\, range})
= K^{{\rm (res)}} \, \f{p_T^3\, d\sigma^{\rm (NLO)}}{dp_T}
({\rm \eta\, in\, exp.\, range})\, ,
\end{equation}
where $K^{{\rm (res)}}$ is as defined in Eq.~(\ref{eq:kres}) in terms
of cross sections integrated over the full region of rapidity. In other
words, we rescale the matched resummed result by the ratio of
NLO cross sections integrated over the experimentally relevant
rapidity region or over all $\eta$, respectively.~\cite{sv}

\begin {figure}[t]
\begin{center}
\includegraphics[width = 3.5in]{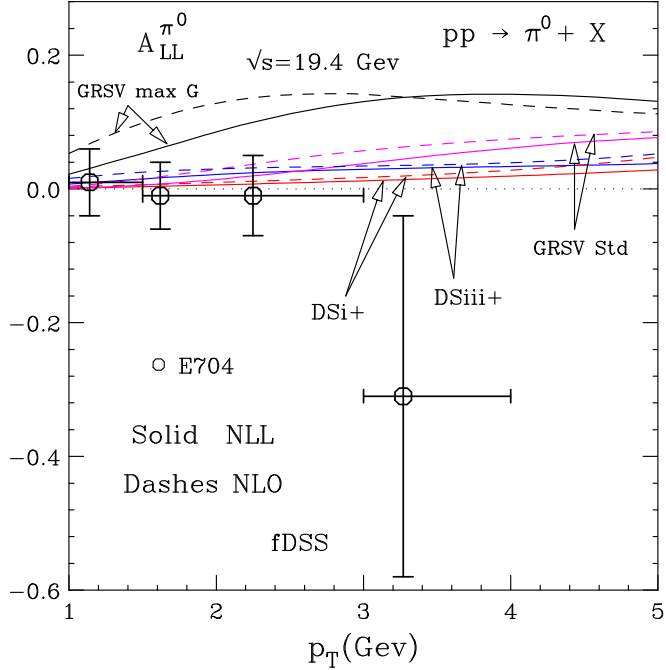}  
\end{center}
\caption{{\it Results for the double-spin asymmetry $A^{\pi}_{LL}$ 
at NLO and for the NLL resummed case for various sets of polarized
parton distributions, at $\sqrt{S}=19.4$~GeV. 
We also show the experimental data of~\cite{experdata}.}
\label{experimdata}}
\end{figure}

Figure~\ref{experimdata} shows our results for the spin asymmetry
$A_{LL}^{\pi}$ at $\sqrt{S}=19.4$~GeV, for the NLO and NLL resummed
cases, defined as in Eq.~(\ref{alldef}), averaged over the pion's 
Feynman-$x_F=x_T \sinh(\eta)$, $|x_F|\leq 0.1$. Again the scales 
have been chosen to be $\mu=p_T$. We also show the data by the Fermilab E704 
experiment~\cite{experdata}. As expected from Fig.~\ref{figres315}, 
$A_{LL}^{\pi}$ generally decreases significantly from NLO to NLL. 
After NLL resummation, even a set with a very large $\Delta g$, 
such as the GRSV ``maximal'' scenario (which is now already ruled out 
by other measurements~\cite{phenix,star,lepton}) shows rough agreement
with the data, given the rather large experimental uncertainties. 
It is interesting to note that similar results were found in~\cite{weber} 
on the basis of LO studies invoking ``intrinsic-$k_T$'' effects.

\begin {figure}[t]
\begin{center}
\includegraphics[width = 5.5in]{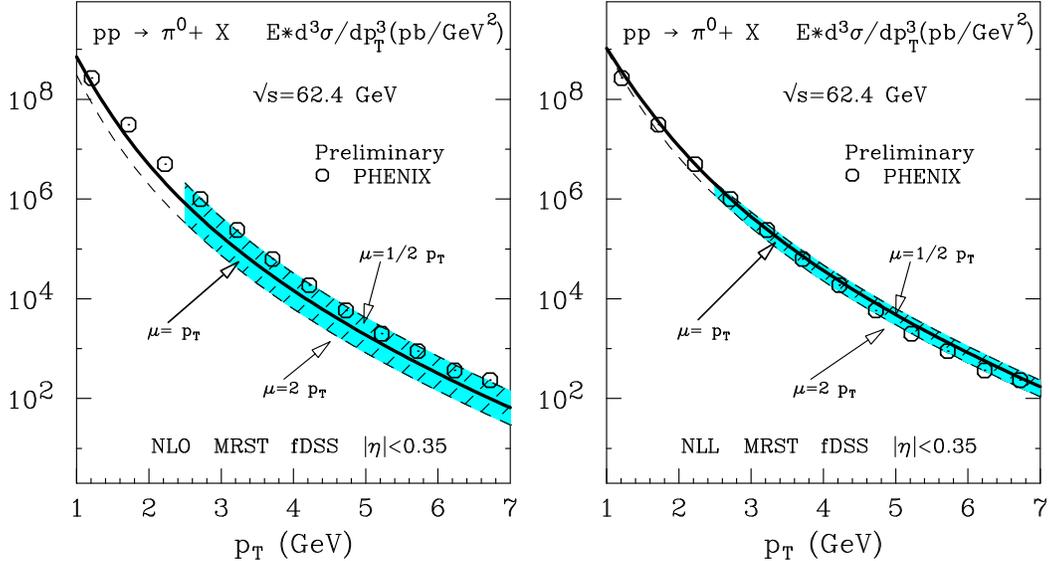}  
\end{center}
\caption{{\it Invariant cross section for $pp\to \pi^0X$ at 
$\sqrt{S}=62.4$ GeV at NLO and for the NLL resummed case. We also
show the preliminary Phenix data~\cite{Tann}.}
\label{fig4}}
\end{figure}

We now return to the case of $pp$ scattering at RHIC at 
$\sqrt{S}=62.4$~GeV. Recently, first preliminary data for the spin-averaged 
high-$p_T$ pion cross section as well as for the spin asymmetry 
$A_{LL}^{\pi}$ were reported by the Phenix collaboration~\cite{Tann}. 
The data cover the pseudo-rapidity region $|\eta|\leq 0.35$. Fig.~\ref{fig4} 
compares our NLO and NLL resummed results for the spin-averaged cross
section to the Phenix data. We use the 
scales $\mu=\zeta p_T$ with $\zeta =1/2, 1, 2$. It is interesting
to see that the data lie at the upper end of the rather wide NLO 
scale band, whereas the resummed predictions have a smaller scale 
dependence and tend to describe the data rather well with scale $\mu=p_T$.
We remind the reader that at $\sqrt{S}=200$~GeV the RHIC data are very well
described by NLO with scale $\mu=p_T$~\cite{phenix}, while in the 
fixed-target regime resummation effects were found to be very
significant (see~\cite{DW1} and Fig.~\ref{figres315} above). We 
interpret all these features as indicating that threshold logarithms 
start to become relevant at $\sqrt{S}=62.4$~GeV, which is ``half way''
between the typical fixed-target regime and RHIC's $200$~GeV. 
\begin {figure}[h]
\begin{center}
\includegraphics[width = 3.5in]{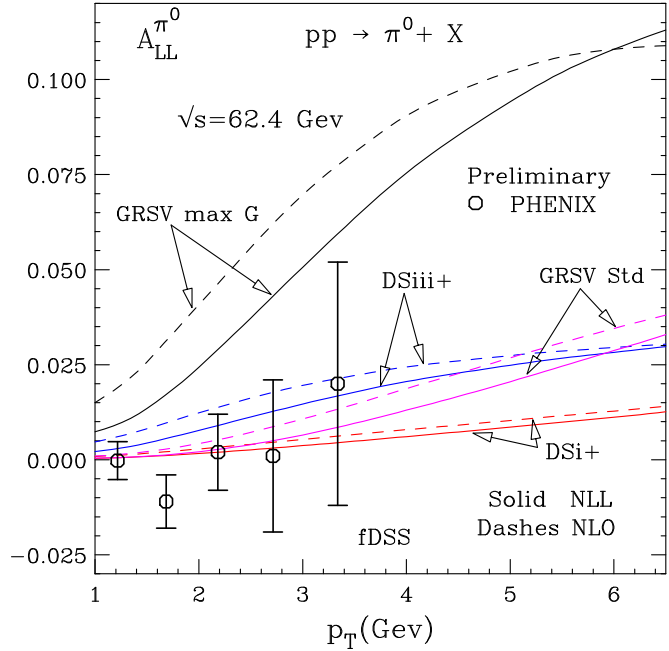}  
\end{center}
\caption{{\it Results for the double-spin asymmetry $A^{\pi}_{LL}$ 
at NLO and for the NLL resummed case for various sets of polarized
parton distributions, at $\sqrt{S}=62.4$~GeV. 
We also show the preliminary experimental data of~\cite{experdata1}.}
\label{fig5}}
\end{figure}
Encouraged by the results in Fig.~\ref{fig4}, 
we show in Fig.~\ref{fig5} our NLO and NLL 
results for the spin asymmetry $A_{LL}^{\pi}$ at $\sqrt{S}=62.4$~GeV, 
along with the Phenix data~\cite{experdata1}. As can be 
seen in Fig.~\ref{fig5}, sets with a large gluon polarization, 
like ``GRSV max G'', show a clear disagreement with the preliminary data.  
We observe that there is again a decrease of $A_{LL}^{\pi}$ when going
from NLO to NLL, but that the resummation effects are somewhat
smaller than what we found at $\sqrt{S}=19.4$~GeV.
We have explicity checked that very similar results for the asymmetries are obtained when implementing other sets of fragmentation functions, like those from~\cite{Kretzer,  Kniehl}. 
\section{Conclusions  \label{sec5}}

We have studied in this paper the NLL resummation of threshold
logarithms in the partonic cross sections relevant for the process
$pp\to h X$ at high transverse momentum of the hadron $h$,
when the initial protons are longitudinally polarized. We have found that,
like for the spin-averaged case~\cite{DW1}, the resummation effects
are large for the spin-dependent cross section in the typical
fixed-target regime at $\sqrt{S}\sim 20$~GeV. The spin asymmetry
$A_{LL}^{\pi}$ is significantly reduced by resummation. A phenomenological
consequence is that the Fermilab E704 data~\cite{experdata} are
compatible with essentially all currently sets of 
spin-dependent parton distribution functions, among them sets with
a rather large gluon polarization. 

We have also applied the resummation to the case $\sqrt{S}\sim 62.4$~GeV,
at which recently preliminary data from RHIC have become available for both 
the spin-averaged cross section and the double-spin 
asymmetry~\cite{Tann, experdata1}. For this case sets with a very 
large gluon polarization in the range $0.05\lesssim x\lesssim 0.2$ are 
ruled out and distributions with a moderate to small gluon polarization
are favored by the data. We find that resummation tends to 
lead to an improved description of the cross section. Its effect
on the spin asymmetry is rather modest. We remind the reader that the 
threshold resummation is somewhat less reliable in this kinematic regime
since one is further away from threshold than for the fixed-target
case, so that subleading perturbative corrections may be more relevant.
It will be desirable in the future to further improve the resummed 
calculation at $\sqrt{S}=62.4$~GeV by including terms that are subleading 
near threshold. More reliable conclusions regarding the effects of 
threshold resummation on $A_{LL}^{\pi}$ at this energy should then
become possible.

\section*{Acknowledgments}
The work of D.dF has been partially supported by Conicet,
 UBACyT and ANPCyT.
W.V.\ is grateful to the 
U.S.\ Department of Energy (contract number DE-AC02-98CH10886) for
providing the facilities essential for the completion of his work.
 The work of F.W. has been supported by UBACyT.

\newpage


\appendix

\section*{Appendix}

\mysection{Results for the various subprocesses}

In this appendix we compile the moment-space expressions
for the spin-dependent Born cross sections for the various partonic 
subprocesses, and the polarized process-dependent coefficients
$\Delta C_{ab\to cd}^{(1)}$, $\Delta G_{I\, ab\to cd}$ contributing
to Eq.~(\ref{eq:res}). As we mentioned
in the main text, the coefficients $D_{I\, ab\to cd}$ are the same 
as in the unpolarized case; they have been given in Ref.~\cite{DW1}. 
Since the $\Delta C_{ab\to cd}^{(1)}$  have rather lengthy expressions,
we only give their numerical values for $N_f=5$ and the
factorization and renormalization scales set to $\mu=Q$. 
In all expressions below, $C_A=3$ and $C_F=(C_A^2-1)/2C_A=4/3$.

\begin{description}
\item[$qq'\to qq'$:]
\begin{eqnarray}
&&\Delta \hat{\sigma}^{{\rm (Born)}}_{qq'\to qq'} (N) = \alpha_s^2
\frac{\pi C_F}{3C_A} \left(
3 N^2 + 5 N \right) B\left(N, \frac{5}{2}\right) \; , \nn \\
&& \Delta G_{1\, qq'\to qq'}=  G_{1\, qq'\to qq'}, \,\,\,\,\, 
\Delta G_{2\, qq'\to qq'}=G_{2\, qq'\to qq'}\, ,\nn \\
&&\Delta C^{(1)}_{1\, qq'\to qq'}=17.9311 \,\,(N_f=5)\; .
\end{eqnarray}
\item[$q\bar{q'}\to q\bar{q'}$:]
\begin{eqnarray}
&& \Delta \hat{\sigma}^{{\rm (Born)}}_{q\bar{q'}\to q\bar{q'}} (N) = \alpha_s^2
\frac{\pi C_F}{3C_A} \left(
3 N^2 + 5 N\right) B\left(N, \frac{5}{2}\right) \; , \nn \\ 
&& \Delta G_{1\, q\bar{q'}\to q\bar{q'}}=  
G_{1\, q\bar{q'}\to q\bar{q'}},\,\,\,\,\, 
\Delta G_{2\, q\bar{q'}\to q\bar{q'}}=G_{2\, 
q\bar{q'}\to q\bar{q'}}\, ,\nn \\
&& \Delta C^{(1)}_{1\, q\bar{q'}\to q\bar{q'}}=20.7021 \,\,(N_f=5)\; .
\end{eqnarray}
\item[$q\bar{q}\to q'\bar{q'}$:]
\begin{eqnarray}
\Delta \hat{\sigma}^{{\rm (Born)}}_{q\bar{q}\to q'\bar{q'}} (N) 
&=& -\hat{\sigma}^{{\rm (Born)}}_{q\bar{q}\to q'\bar{q'}} (N) , \nn \\
\Delta G^{(1)}_{1\, q\bar{q}\to q'\bar{q'}}=G^{(1)}_{1\, 
q\bar{q}\to q'\bar{q'}}, && \Delta C^{(1)}_{1\, 
q\bar{q}\to q'\bar{q'}}=C^{(1)}_{1\, q\bar{q}\to q'\bar{q'}}. 
\end{eqnarray}
\item[$qq\to qq$:]
\begin{eqnarray}
&& \Delta \hat{\sigma}^{{\rm (Born)}}_{qq\to qq} (N) =
\alpha_s^2 \frac{2\pi C_F}{3C_A^2} \left(C_A (3 N^2+5N) -
2 N (3+2N)\right)B\left(N, \frac{5}{2}\right) \; , \nn \\
&& \Delta G_{1\, qq\to qq}=7/5\,,\,\,\,\,\, \Delta G_{2\, 
qq\to qq}=-2/5\, ,\,\,\,\,\,  \Delta C^{(1)}_{1\, qq\to qq}=
14.5364 \,\,(N_f=5)\; .
\end{eqnarray}
\item[$q\bar{q}\to q\bar{q}$:]
\begin{eqnarray}
&&\Delta \hat{\sigma}^{{\rm (Born)}}_{q\bar{q}\to q\bar{q}} (N) =
\alpha_s^2 \frac{\pi C_F}{15C_A^2} N \left(C_A (2+N)(11+5N) - 
(N+3)(5+2N)\right)B\left(N, \frac{7}{2}\right) \; , \nn \\
&& \Delta G_{1\,q\bar{q}\to q\bar{q}}=-3/13\,,\,\,\,\,\, \Delta G_{2\,
q\bar{q}\to q\bar{q}}=16/13 \,, \,\,\,\,\, \Delta C^{(1)}_{1\, 
q\bar{q}\to q\bar{q}}=24.2465 \,\,(N_f=5)\; .
\end{eqnarray}
\item[$q\bar{q}\to gg$:]
\begin{eqnarray}
&&\Delta \hat{\sigma}^{{\rm (Born)}}_{q\bar{q}\to gg} (N) = 
-\hat{\sigma}^{{\rm (Born)}}_{q\bar{q}\to gg} (N) , \qquad 
\Delta C^{(1)}_{1\, q\bar{q}\to gg}= C^{(1)}_{1\, q\bar{q}\to gg},\nn \\ 
&& \Delta G_{1\, q\bar{q}\to gg}=  G_{1\, q\bar{q}\to gg}, 
\,\,\,\,\, \Delta G_{2\, q\bar{q}\to gg}=G_{2\, q\bar{q}\to gg}\, .
\end{eqnarray}
\item[$qg\to qg$:]
\begin{eqnarray}
&& \Delta \hat{\sigma}^{{\rm (Born)}}_{qg\to qg} (N) =
\alpha_s^2 \frac{\pi}{6C_A} \left(C_F + 2C_A \right) (3N^2+5N) B\left(N, 
\frac{5}{2}\right) \; , \nn \\
&& \Delta G_{1\, qg\to qg}=  G_{1\, qg\to qg}, \,\,\, \Delta 
G_{2\, qg\to qg}=G_{2\, qg\to qg}, \,\,\, \Delta G_{3\, qg\to qg}=
G_{3\, qg\to qg}\, ,\nn \\
&& \Delta C^{(1)}_{1\, qg\to qg}=14.2048 \,\,(N_f=5)\; .
\end{eqnarray}
\item[$qg\to gq$:]
\begin{eqnarray}
&& \Delta \hat{\sigma}^{{\rm (Born)}}_{qg\to gq} (N) =
\alpha_s^2 \frac{\pi}{6C_A} \left(C_F + 2C_A \right) (3N^2+5N)B\left(N, 
\frac{5}{2}\right) \; , \nn \\
&& \Delta G_{1\, qg\to gq}=  G_{1\, qg\to gq}, \,\,\, 
\Delta G_{2\, qg\to gq}=G_{2\, qg\to gq}, \,\,\, 
\Delta G_{3\, qg\to gq}=G_{3\, qg\to gq}\, ,\nn \\
&&\Delta C^{(1)}_{1\, qg\to gq}=21.2354  \,\,(N_f=5)\; .
\end{eqnarray}
\item[$gg\to gg$:]
\begin{eqnarray}
&&\Delta \hat{\sigma}^{{\rm (Born)}}_{gg\to gg} (N) =
\alpha_s^2 \frac{\pi C_A}{15C_F} \left(21 N^3 +89 N^2 + 92 N \right) 
B\left(N, \frac{7}{2}\right) \; , \nn \\
&& \Delta G_{1\, gg\to gg}=  G_{1\, gg\to gg}, \,\,\, \Delta 
G_{2\, gg\to gg}=G_{2\, gg\to gg}, \,\,\, \Delta G_{3\, gg\to gg}=
G_{3\, gg\to gg}\, ,\nn \\
&&\Delta C^{(1)}_{1\, gg\to gg}=20.3233 \,\,(N_f=5)\; .
\end{eqnarray}
\item[$gg\to q\bar{q}$:]
\begin{eqnarray}
&& \Delta \hat{\sigma}^{{\rm (Born)}}_{gg\to q\bar{q}} (N)=
-\hat{\sigma}^{{\rm (Born)}}_{gg\to q\bar{q}} (N) ,\qquad 
\Delta C^{(1)}_{1\, gg\to q\bar{q}} = C^{(1)}_{1\, gg\to q\bar{q}}, \nn \\
&& \Delta G_{1\, gg\to q\bar{q}}=  G_{1\, gg\to q\bar{q}}, 
\,\,\,\,\, \Delta G_{2\,gg\to q\bar{q}}=G_{2\, gg\to q\bar{q}}\, .
\end{eqnarray}
\end{description}

In the above expressions, $B(a,b)$ is the Euler Beta-function.

\mysection{Spin-dependent color-connected Born cross sections}

In this appendix we compile the color-connected Born cross sections
that we need for our study. Our choices for the color bases follow 
precisely~\cite{KOS}. For a given color basis, the color-connected Born 
cross sections will appear as a matrix that we shall refer to as
``hard matrix''. We will not repeat the expressions for the soft 
matrices $S$, and the anomalous dimension matrices $\Gamma$ in these
bases, which may all be found in~\cite{KOS}. Like~\cite{KOS}, we present 
our results for arbitrary partonic rapidity, even though for our actual 
study of the rapidity-integrated cross section we only need the case
$\hat\eta=0$. For each partonic reaction $ab\to cd$, we define
the Mandelstam variables $s=(p_a+p_b)^2$, $t=(p_a-p_c)^2$, 
$u=(p_a-p_d)^2$. $t,u$ are functions of $\hat\eta$. 
In all expressions below, $N_c=3$ and $C_F=(N_c^2-1)/2N_c=4/3$.

We begin with the quark-antiquark annihilation processes. Depending on 
the quark flavor, there are three different quark-antiquark subprocesses 
to consider, $q_j {\bar q}_j \rightarrow q_j {\bar q}_j $ , $q_j 
{\bar q}_j \rightarrow q_k {\bar q}_k $ and $q_j {\bar q}_k 
\rightarrow q_j {\bar q}_k $. Each of these have their own hard
matrix elements:

$q_j {\bar q}_j \rightarrow q_j {\bar q}_j $
\beqa
\Delta H_{11}^{q_j {\bar q_j}\rightarrow q_j {\bar q_j}} &=& 
- H_{11}^{q_j {\bar q_j}\rightarrow q_j {\bar q_j}}
= - \alpha_s^2 \frac{2C_F^2}{N_c^4} \frac{(t^2+u^2)}{s^2} \, , \nonumber \\
\Delta H_{12}^{q_j {\bar q_j}\rightarrow q_j {\bar q_j}} &=& 
-  H_{12}^{q_j {\bar q_j}\rightarrow q_j {\bar q_j}}
= - \alpha_s^2 \frac{2C_F}{N_c^3} \left[-\frac{(t^2+u^2)}{N_c s^2}
 +\frac{u^2}{st}\right] \, ,\nonumber \\
\Delta H_{22}^{q_j {\bar q_j}\rightarrow q_j {\bar q_j}}
&=& \alpha_s^2 \frac{1}{N_c^2} \left[- \frac{2}{N_c^2}\frac{(t^2+u^2)}{s^2}
+2\frac{(s^2-u^2)}{t^2}+\frac{4}{N_c}\frac{u^2}{st}\right] \, .
\eeqa
where $H_{i j}^{ab\to cd}$ refers in each case to the corresponding 
hard matrix for the unpolarized case given in~\cite{KOS}.
 
$q_j {\bar q}_j \rightarrow q_k {\bar q}_k $

Here one has simply:
\beq
\Delta H^{q_j {\bar q_j}\rightarrow q_k {\bar q_k}} = - H^{q_j 
{\bar q_j}\rightarrow q_k {\bar q_k}}  \; .
\eeq

$q_j {\bar q}_k \rightarrow q_j {\bar q}_k $

The polarized hard matrix at lowest order can be expressed as,

\beq
\Delta H^{q_j {\bar q_k}\rightarrow q_j {\bar q_k}}=\alpha_s^2 \left[
                \begin{array}{cc}
                 0 & 0  
\vspace{2mm} \\
                 0 & 2(s^2-u^2)/(N_c^2 t^2)
               \end{array} \right] \, .
\eeq

There are only two different quark-quark processes to consider, 
depending on the quark flavors:

$q_j q_k \rightarrow q_j q_k $

Here the result is the same as for the process 
$q_j \bar{q}_k \rightarrow q_j \bar{q}_k$, but with the 
color basis inverted (equivalent to an interchange of the entries
in the first and second rows and columns, $c_1 \to c_2$).

$q_j q_j \rightarrow q_j q_j $

Here, 
\beqa
\Delta H_{11}^{q_j q_j \rightarrow q_j q_j}
&=& \alpha_s^2 \frac{2}{N_c^2} \left[\frac{(s^2-u^2)}{t^2}
+\frac{1}{N_c^2}\frac{(s^2-t^2)}{u^2}-\frac{2}{N_c}\frac{s^2}{tu}\right] \, ,
\nonumber \\
\Delta H_{12}^{q_j q_j \rightarrow q_j q_j}
&=& \alpha_s^2 \frac{2C_F}{N_c^4} \left[N_c\frac{s^2}{tu}
-\frac{(s^2-t^2)}{u^2}\right] \, ,
\nonumber \\
\Delta H_{22}^{q_j q_j \rightarrow q_j q_j}
&=& \alpha_s^2 \frac{2C_F^2}{N_c^4} \frac{(s^2-t^2)}{u^2} \, .
\eeqa

\newpage
$q_j q_k \rightarrow q_j q_k $

In the spin-dependent case one has:
\beq
\Delta H^{q_j q_k \rightarrow q_j q_k}= \alpha_s^2 \left[
                \begin{array}{cc}
                 2(s^2-u^2)/(N_c^2 t^2) & 0  
\vspace{2mm} \\
                 0 & 0
               \end{array} \right] \, .
\eeq

$q {\bar q} \rightarrow  gg$

Here, 
\beq
\Delta H^{q {\bar q} \rightarrow gg}= -H^{q {\bar q} \rightarrow gg} \; .
\eeq

$gg \rightarrow  q {\bar q} $

Again, the polarized coefficients are the negatives of the unpolarized ones:
\beq
\Delta H^{gg \rightarrow q {\bar q}}= -H^{gg \rightarrow q {\bar q}} \; .
\eeq

${q g \rightarrow  q g}$

Here, 
\beq
\Delta H_{ij}^{qg \rightarrow qg}= \frac{s^2-u^2}{s^2+u^2} 
H_{ij}^{qg \rightarrow qg} \; .
\eeq

Finally, we consider gluon-gluon scattering. For simplicity, we set
$N_c=3$ explicitly here.

$g g \rightarrow  g g$
\beq
\Delta H_{ij}^{gg \rightarrow gg}= \frac{s^4-t^4-u^4}{s^4+t^4+u^4} 
H_{ij}^{qg \rightarrow qg} \; .
\eeq



\begin{thebibliography}{99}

\bibitem{grsv}
  M.~Gl\"{u}ck, E.~Reya, M.~Stratmann and W.~Vogelsang,
  Phys.\ Rev.\  D {\bf 63}, 094005 (2001) [arXiv:hep-ph/0011215].

\bibitem{DS} D.~de Florian and R.~Sassot,
  Phys.\ Rev.\ D {\bf 62}, 094025 (2000)
  [arXiv:hep-ph/0007068].

\bibitem{deltag}  
 J.~Bl\"{u}mlein and H.~B\"{o}ttcher,
  Nucl.\ Phys.\  B {\bf 636}, 225 (2002)
  [arXiv:hep-ph/0203155]; 
  E.~Leader, A.~V.~Sidorov and D.~B.~Stamenov,
  Phys.\ Rev.\  D {\bf 73}, 034023 (2006)
  [arXiv:hep-ph/0512114]; 
  M.~Hirai, S.~Kumano and N.~Saito,
  Phys.\ Rev.\  D {\bf 74}, 014015 (2006)
  [arXiv:hep-ph/0603213]; 
 D.~de Florian, G.~A.~Navarro and R.~Sassot,
  Phys.\ Rev.\  D {\bf 71}, 094018 (2005) [arXiv:hep-ph/0504155].

\bibitem{Stratmann:2007hp}
  M.~Stratmann and W.~Vogelsang,
  arXiv:hep-ph/0702083.

\bibitem{phenix} A.~Adare {\it et al.} [PHENIX Collaboration],
arXiv:0704.3599 [hep-ex]; K.~Boyle, 
talk presented at the 2007 RHIC \& AGS Annual Users' Meeting, BNL, 
June 18-22, 2007.

\bibitem{star} B.~I.~Abelev {\it et al.}  [STAR Collaboration],
Phys.\ Rev.\ Lett.\  {\bf 97}, 252001 (2006)  [arXiv:hep-ex/0608030]; 
S.\ Vigdor [STAR Collaboration], ``New Results from RHIC on the Spin Structure 
of the Proton'', plenary talk presented at the 
APS April meeting, April 14-17, 2007, Jacksonville, Florida;
M.~Sarsour, 
talk presented at the 2007 RHIC \& AGS Annual Users' Meeting, BNL, 
June 18-22, 2007.

\bibitem{lepton} B.~Adeva {\it et al.} [Spin Muon Collabroation],
Phys.\ Rev.\ {\bf D70}, 012002 (2004); E.S.~Ageev {\it et al.}  
[COMPASS Collaboration], Phys.\ Lett. {\bf B633}, 25 (2006);
A.~Airapetian {\it et al.} [HERMES Collaboration],
Phys.\ Rev.\ Lett. {\bf 84}, 2584 (2000); P.~Liebing,
talk presented at the ``17th International Spin Physics Symposium 
(Spin 2006)'', Kyoto, Japan, October 2-7, 2006.

 \bibitem{experdata}
  D.~L.~Adams {\it et al.}  [E581 \& E704 Collaborations],
  Phys.\ Lett.\ B {\bf 261}, 197 (1991).

\bibitem{aur} P.~Aurenche, M.~Fontannaz, J.~P.~Guillet, B.~A.~Kniehl,
and M.~Werlen, Eur.\ Phys.\ J.\ C {\bf 13}, 347 (2000) 
[arXiv:hep-ph/9910252]; 
U.~Baur {\it et al.}, arXiv:hep-ph/0005226; 
C.~Bourrely and J.~Soffer,
Eur.\ Phys.\ J.\ C {\bf 36}, 371 (2004) [arXiv:hep-ph/0311110].

\bibitem{weber} W.~Vogelsang and A.~Weber,
Phys.\ Rev.\  D {\bf 45}, 4069 (1992).

\bibitem{DW1}
 D.~de Florian and W.~Vogelsang,
  Phys.\ Rev.\ D {\bf 71}, 114004 (2005)
  [arXiv:hep-ph/0501258].

\bibitem{dyresum} G.~Sterman, Nucl.\ Phys.\ B {\bf 281}, 310 (1987);
S.~Catani and L.~Trentadue, Nucl.\ Phys.\ B {\bf 327}, 323 (1989);
Nucl.\ Phys.\ B {\bf 353}, 183 (1991).

\bibitem{KOS} N.~Kidonakis and G.~Sterman,
Nucl.\ Phys.\ B {\bf 505}, 321 (1997) [arXiv:hep-ph/9705234]; 
N.~Kidonakis, G.~Oderda and G.~Sterman,
Nucl.\ Phys.\ B {\bf 525}, 299 (1998) [arXiv:hep-ph/9801268];
Nucl.\ Phys.\ B {\bf 531}, 365 (1998) [arXiv:hep-ph/9803241];
N.~Kidonakis and J.~F.~Owens, Phys.\ Rev.\ D {\bf 63},
054019 (2001) [arXiv:hep-ph/0007268].

\bibitem{Bon} R.~Bonciani, S.~Catani, M.~L.~Mangano and P.~Nason,
Phys.\ Lett.\ B {\bf 575}, 268 (2003) [arXiv:hep-ph/0307035]

\bibitem{bpr} S.~J.~Brodsky, H.~J.~Pirner and J.~Raufeisen,
Phys.\ Lett.\  B {\bf 637}, 58 (2006) [arXiv:hep-ph/0510315].

\bibitem{Tann} M.~J.~Tannenbaum (PHENIX Collaboration), arXiv:0707.1679.

 \bibitem{experdata1}
K.~Aoki  [PHENIX Collaboration],
talk presented at the ``17th International Spin Physics Symposium 
(Spin 2006)'', Kyoto, Japan, October 2-7, 2006,
  arXiv:0704.1369 [hep-ex]. These data are preliminary.

\bibitem{FT}
   S.~B.~Libby and G.~Sterman,
  Phys.\ Rev.\  D {\bf 18}, 3252 (1978); 
   R.~K.~Ellis, H.~Georgi, M.~Machacek, H.~D.~Politzer and G.~G.~Ross,
  Phys.\ Lett.\  B {\bf 78}, 281 (1978);
  D.~Amati, R.~Petronzio and G.~Veneziano,
  Nucl.\ Phys.\  B {\bf 146}, 29 (1978);
  Nucl.\ Phys.\  B {\bf 140}, 54 (1978); 
  G.~Curci, W.~Furmanski and R.~Petronzio,
  Nucl.\ Phys.\  B {\bf 175}, 27 (1980); 
    J.~C.~Collins, D.~E.~Soper and G.~Sterman,
  Phys.\ Lett.\  B {\bf 134}, 263 (1984);
  Nucl.\ Phys.\  B {\bf 261}, 104 (1985); 
  J.~C.~Collins,
  Nucl.\ Phys.\  B {\bf 394}, 169 (1993).

\bibitem{CMN} S.~Catani, M.~L.~Mangano and P.~Nason, JHEP {\bf 9807}, 
024 (1998) [arXiv:hep-ph/9806484].

\bibitem{cc} M.~Cacciari and S.~Catani, Nucl.\ Phys.\ B {\bf 617},
253 (2001) [arXiv:hep-ph/0107138].

\bibitem{KT} J.~Kodaira and L.~Trentadue,
Phys.\ Lett.\ B {\bf 112}, 66 (1982); Phys.\ Lett.\ B {\bf 123},
335 (1983); S.~Catani, E.~D'Emilio and L.~Trentadue,
Phys.\ Lett.\ B {\bf 211}, 335 (1988).


\bibitem{Jager}
B.\ J\"{a}ger, A.\ Sch\"{a}fer, M.\ Stratmann,
and W.\ Vogelsang, Phys. Rev. {\bf D67}, 054005 (2003)
[arXiv:hep-ph/0211007].

\bibitem{Catani:1996yz} S.~Catani, M.~L.~Mangano, P.~Nason
and L.~Trentadue, Nucl.\ Phys.\ B {\bf 478}, 273 (1996)
[arXiv:hep-ph/9604351].


\bibitem{DdF}
D. de Florian, Phys. Rev. D {\bf 67} (2003) 054004 [arXiv:hep-ph/0210442].

\bibitem{mrst}
  A.~D.~Martin, R.~G.~Roberts, W.~J.~Stirling and R.~S.~Thorne,
  Eur.\ Phys.\ J.\ C {\bf 35}, 325 (2004)
  [arXiv:hep-ph/0308087].

\bibitem{deFlorian:2007aj}
  D.~de Florian, R.~Sassot and M.~Stratmann,
  arXiv:hep-ph/0703242.

\bibitem{sv} G.~Sterman and W.~Vogelsang,
JHEP {\bf 0102}, 016 (2001) [arXiv:hep-ph/0011289].

\bibitem{Kretzer}
  S.~Kretzer,
  Phys.\ Rev.\  D {\bf 62}, 054001 (2000)
  [arXiv:hep-ph/0003177].

\bibitem{Kniehl}
  B.~A.~Kniehl, G.~Kramer and B.~Potter,
  Nucl.\ Phys.\  B {\bf 582}, 514 (2000)
  [arXiv:hep-ph/0010289].

\end{thebibliography}
\end{document}